\documentstyle[prb,floats,aps,twocolumn,psfig]{revtex}

\begin{document}

\draft 
\tightenlines
\widetext

\title{Correlated ab-initio calculations for ground-state properties
of II-VI semiconductors}

\author{Martin Albrecht and Beate Paulus} 
\address{Max-Planck-Institut f\"ur
Physik komplexer Systeme, Bayreuther Stra\ss e 40, D-01187 Dresden, Germany}
\author{Hermann Stoll} \address{Institut f\"ur Theoretische Chemie,
Universit\"at Stuttgart, D-70550 Stuttgart, Germany}

\maketitle

\begin{abstract}
Correlated {\em ab-initio} ground-state calculations, using relativistic
energy-consistent pseudopotentials, are performed for six 
II-VI semiconductors. Valence ($ns,np$) correlations are
evaluated using the coupled cluster approach with
single and double excitations.
An incremental scheme is applied based on correlation contributions of
localized bond orbitals and of pairs and triples of such bonds. 
In view of the high polarity of the bonds in II-VI compounds,
we examine both, ionic and covalent embedding schemes for the
calculation of individual bond increments. Also, a
partitioning of the correlation energy according to local ionic increments is
tested. Core-valence ($nsp,(n-1)d$) correlation effects are taken into
account via a core-polarization potential. Combining the results at the
correlated level with corresponding Hartree-Fock data we recover about
94\% of the experimental cohesive energies; lattice constants are accurate 
to $\sim$1\%; bulk moduli are on average 10\% too large compared with 
experiment.     
\end{abstract} 

\narrowtext

\section{Introduction}
{\em Ab initio} calculations for solids are mostly performed nowadays 
formally within
a one-particle picture using density-functional theory (DFT), i.e.\
describing the non-local exchange together with many-particle
correlation effects within the local-density approximation
(LDA)\cite{jones89} or within more sophisticated
generalized-gradient approximations (GGA)\cite{perdew95}.
As an alternative,
it is possible at the one-particle level to calculate
the non-local exchange of a solid exactly, using a periodic Hartree-Fock (HF)
scheme\cite{pisani88}.\\
However, for a proper microscopic treatment of electron correlations
it is necessary to go beyond
the one-particle framework and to deal with many-body wave functions.
In the case of finite systems, accurate quantum-chemical methods, of the 
configuration interaction (CI) or coupled-cluster (CC) type, have been
developed for the determination of many-body wave functions and
the corresponding correlation energies. In the case of solids, 
the latter methods
are not directly applicable, but it is often possible, to a very 
good approximation,
to cast the correlation energy of the infinite system into a rapidly 
converging expansion
in terms of increments from localized orbital subsystems\cite{stoll92}.
Due to the local character of the correlation hole
these increments may be determined for finite fragments of the solid and, as a
consequence, the correlation energy of the solid becomes
accessible to a quantum-chemical treatment.\\
The method of local increments has been applied to elementary\cite{paulus95}
and III-V semiconductors\cite{paulus96}, with an expansion in 
terms of correlation contributions related to localized bond orbitals 
and to pairs and triples
of such bonds. On the other hand, the same method has also successfully been
used for ionic compounds such as MgO, CaO, and NiO\cite{doll95}; there, 
the local
increments refer to groups of atomic orbitals which can be assigned to 
given anionic or cationic centers.
In an attempt to bridge the 
gap between covalent and ionic
solids we now investigate the use of incremental schemes for 
correlation effects in
II-VI semiconductors.\\
In Section II, we present the results of Hartree-Fock self-consistent
field (SCF) calculations for the solids just mentioned. The inclusion 
of correlations is discussed
in Section III; we first focus on the various
correlation effects which have to be considered, and then elaborate on
embedding procedures and the computational details of the method.
Results follow in  Section IV. Finally, pertinent conclusions are given 
in Section V.

\section{Hartree-Fock calculations}

A starting point for treating many-body correlation effects in solids are
reliable Hartree-Fock self-consistent field results for the infinite
system. We performed HF ground-state calculations for six 
II-VI semiconductors, i.e., ZnS, ZnSe and ZnTe and the corresponding
Cd compounds in the zinc-blende structure, using the 
program package {\sc Crystal92}\cite{crystal92}.\\ 
The large number of electrons 
and the relativistic effects occurring in these materials can efficiently 
be handled
by means of pseudopotentials.
For the group VI elements we use the scalar-relativistic 
energy-consistent pseudopotentials
of Bergner et al., together with the corresponding 
$(4s5p)/[3s3p]$ atomic valence basis sets\cite{bergner93}.  
For their application in the {\sc Crystal} calculations the basis sets 
have to be modified, however:
very diffuse exponents which are necessary to properly describe the 
tails of the free-atom wavefunctions cause numerical problems in {\sc Crystal}.
In the solid the basis functions of
neighbouring atoms take over their role due to close-packing.
Therefore we leave out the most diffuse $p$ 
exponent and use a reduced atomic $(4s4p)/[3s3p]$ basis set
supplemented by a $d$ polarization function, which has been  optimized 
for the Zn-compounds in the solid ($d_{\rm S}$=0.45, $d_{\rm Se}$=0.35,
$d_{\rm Te}$=0.25).\\
For the group IIb elements Zn and Cd we use two different sets of 
pseudopotentials:
first 20-valence-electron pseudopotentials,
where the outer-core  $(n-1)d$ shell and the corresponding $(n-1)s,p$ 
shells are
explicitly treated together with the $ns,p$ valence 
electrons\cite{dolg86,andrae90}.
Again,
optimized $(8s7p6d)/[6s5p3d]$ basis sets are available (which we use for the
calculation of the free atoms), but a reoptimization is needed for the solid.
Leaving out the most diffuse $s$ and $p$ exponents of the atomic basis sets
(for Cd the outer $d$ exponent, too),
we determined, for each angular
momentum independently, the smallest exponent which still leads 
to a stable solution
in the {\sc Crystal} calculation. Starting with these (fixed) outer exponents
we reoptimized the inner ones yielding  $(7s6p6d$ or $7s6p5d)/[5s4p2d]$ 
basis sets (see Table \ref{basis}).\\
Furthermore, we use large-core 2-valence electron pseudopotentials for Zn
\cite{dolgunp}(see Table \ref{pp}) and Cd
\cite{igelmann87} (which 
will also be used at the correlated level, cf.\ Sect.\ 3).
\begin{table}
\caption{\label{pp}The Zn 2-valence-elelctron pseudopotential is represented 
in the 
semilocal form $V_{\rm PP}(r_i)=-\frac{Q}{r_i}+\sum_{l=0}^{l_{\rm max}}
\sum_k A_{lk}{\rm exp}(-\alpha_{lk}r_i^2) P_l$, where $P_l$ is the 
projection operator onto the Hilbert subspace of angular symmetry $l$.
The exponents $\alpha_{lk}$ and the coefficients $A_{lk}$ are listed below.}
\begin{tabular}{c||c|c}
$l$&$\alpha_{lk}$&$A_{lk}$\\
\hline
$s$&1.4988020&18.31672\\
&0.7490050&-3.405011\\
$p$&1.5327700&11.46430\\
&0.7870910&-1.327391\\
$d$&0.7502760&1.583946\\
&0.3747920&0.333476\\
$f$&0.4666990&-0.398428
\end{tabular}
\end{table}
For Zn, a $(4s2p)$ basis set 
is available in the 
literature\cite{dolgunp}. Since in the materials which we are concerned 
with (polarized) bonds
between $sp^3$-hybrids are formed, we 
added two inner $p$ functions optimized in a Hartree-Fock calculation 
for the first excited $s^1p^1$ $^3P$ atomic state. The 
outermost $p$ exponent is subsequently left out
in the {\sc Crystal} calculation; in addition, the two most 
diffuse $s$ functions are reoptimized for the solid, and
a $d$ polarization function is added, yielding a 
$(4s3p1d)/[3s3p1d]$ basis set
(see Table \ref{basis}).
For the Cd pseudopotential\cite{igelmann87}  no basis set exists 
in the literature.
We therefore first optimized a $(5s4p)$ basis in a Hartree-Fock 
calculation
for the first excited $s^1p^1$ $^3P$ atomic state.   
For the solid, the outer $s$ and $p$ 
exponents were neglected, and 
a $d$ exponent was optimized yielding a
$(4s3p1d)/[3s3p1d]$ basis set (see Table \ref{basis}).\\ 
Using the small-core pseudopotentials for Zn and Cd with their corresponding
basis sets we determined SCF ground
state energies of the solids. To estimate the numerical accuracy of the 
{\sc Crystal} results
(for a detailed discussion see Ref.[\onlinecite{crystal92}]),
we performed various test
calculations concerning the computational parameters for ZnS, 
the system with the smallest lattice constant.
The convergence with respect to the `Coulomb overlap' parameter is very good
yielding an error of less than $10^{-5}$ Hartree. The `Coulomb penetration'
parameter which controls the Coulomb series is more critical.
We could not reach any monotonous convergence, the deviation between
the two 'best' parameters pointing to an uncertainty of the order
of $10^{-4}$ Hartree. Concerning the exchange series a similar
behaviour is found. The convergence with respect to the 
`exchange overlap' is very good (error less than $10^{-5}$ Hartree), whereas
the parameters affecting the `exchange penetration' are very critical.
Due to the quite diffuse basis functions we had to use, we could only
reach an accuracy of the order of $10^{-4}$ Hartree. Thus,
for ZnS we end up with an error bar of $\pm 3\cdot
10^{-4}$ Hartree.
The finite $k$-point sampling for the reciprocal-space integration 
causes errors of less than $10^{-5}$ Hartree, which can be neglected in
comparison with the error due to the truncation of the Coulomb and 
exchange series.\\ 
For evaluating cohesive energies, we subtracted the
ground-state energies of the free atoms obtained with the corresponding
atom-optimized basis sets; the
quantum-chemical {\em ab-initio} program system {\sc
Molpro94}\cite{molpro94} was used in these latter calculations.  
The results for SCF cohesive energies
at the experimental lattice constants
\cite{landoltiii17b} applying the small-core pseudopotentials for Zn and Cd
are listed in the first column 
of Table \ref{coh}. A comparison is made to experimental
cohesive energies (last column in Table \ref{coh})
which are corrected by  phonon zero-point energies
$\frac{9}{8}{\rm k_B}\Theta_{\rm D}$ (derived from the Debye
model\cite{farid91}) as well as by atomic spin-orbit
splittings\cite{moore}. It is seen that binding energies at the HF
level are between 60\% and 70\% of the experimental values leaving
room for significant correlation contributions.\\
For determining lattice constants and bulk moduli, we performed a
4th order polynomial fit to the SCF ground-state energies
evaluated within a range of 
$-2\%$ to $+6\%$ of the experimental lattice constant, in half-percent
steps. The deviations between calculated and fitted points are
of about $\pm 1\cdot 10^{-4}$ Hartree. 
\begin{table*}
\caption{\label{basis}Crystal-optimized basis sets for
20-valence-electron and  2-valence-electron
pseudopotentials of Zn and Cd. In parentheses, we list the additional exponents and 
contraction coefficients of the atomic basis used in the calculations of the 
free atoms as well as in the cluster calculations.}
\begin{tabular}{c||c|c||c|c||c|c}
&$s$-exp.&coeff.&$p$-exp.&coeff.&$d$-exp.&coeff.\\
\hline
Zn&27.785554&0.1074375&92.652341&0.0024709&61.208798&0.0220562\\
20-ve&17.520914&-0.1821469&19.771450&-0.0781854&19.141955&0.1188778\\
&10.282040&-0.3099337&4.465187&0.4036449&6.873575&0.3064757\\
&2.755438&1&1.908780&0.525450&2.517245&0.4305467\\
&1.169907&1&0.758772&1&0.858306&0.363375\\
&0.187&1&0.10&1&0.25&1\\
&0.11&1&&&&\\
\hline
Zn&1.572755&0.313862&1.025&1 (-0.076200)&0.235&1\\
2-ve&1.198905&-0.541801&0.26&1 (0.269338)&&\\
&0.20 (0.148856)&1&0.13627&1&&\\
&0.10 (0.051016)&1&(0.0415)&&&\\
\hline
Cd&10.497284&0.4871518&5.130033&-0.5067959&8.890067&-0.0138837\\
20-ve&6.998189&-1.0501142&3.420022&0.6070007&2.964186&0.2810514\\
&4.665459&0.0728619&1.282189&0.6495556&1.231907&0.4924337\\
&1.465187&1&0.543303&0.2159152&0.465127&0.3622932\\
&0.654782&1&0.169429&1&0.15&1\\
&0.355802&1&0.08&1&&\\
&0.08&1&&&&\\
\hline
Cd&1.298843&0.240843&0.581900&1 (-0.192137)&0.203&1\\
2-ve&0.865895&-0.534798&0.337511&1 (0.227533)&&\\
&0.202302&1 (0.232332)&0.095653&1&&\\
&0.090463&1&(0.033746)&&&\\
&(0.037232)&1&&&\\
\end{tabular}
\end{table*}
The lattice constants (listed in Table \ref{latt}) obtained from the
minimum of the polynomial fit are too large,
by up to 4.5\%, as compared to experiment.
When applying different kinds of fits,
to all calculated points as well as to those surrounding the minimum only,
the lattice constants change by at most
$\pm$ 0.01\AA ($\sim$0.2\%). Thus, we conclude that correlation corrections are improtant for
the lattice constants, too.\\
The bulk modulus of cubic structures can be determined according to
\begin{equation}
B=\left( \frac{4}{9a}\frac{\partial^2}{\partial a^2} - \frac{8}{9a^2}
\frac{\partial}{\partial a} \right) E_{{\rm total}} (a),
\end{equation}
where $a$ is the lattice constant. The second term is only zero if we calculate the bulk modulus
at the minimum of the potential curve.
The first and especially the second derivative of the potential curve
are much more affected by the errors of the total energy
than the position of the minimum of the curve. Taking into account all
points in the range of $-2$\% to $+$6\% of the experimental
lattice constant, the total-energy error can be estimated to lead to an uncertainty
of about $\pm$2\% for the bulk modulus; with only
seven points around the minimum this error bar would increase to up to
$\pm$15\%.  We therefore performed a 4th order polynomial fit
to all points when determining $\frac{\partial E}{\partial a}$ and 
$\frac{\partial^2 E}{\partial a^2}$.
Additionally, we tested polynomial fits of other degrees:
we find that 
3rd, 4th or 5th order fits differ by about $\pm$3\% only, whereas 
a quadratic fit would yield bulk moduli smaller by up to 20\%.
We calculated the bulk moduli at the Hartree-Fock lattice
constant and at the experimental one.   
The bulk moduli at the Hartree-Fock lattice constant are too small 
by up to $\sim$30\%, 
compared with the experiment, mainly  
due to the drastic overestimation of the lattice constants.
The Hartree-Fock bulk moduli calculated at the experimental lattice 
constant, on the other hand, are by up to 30\% higher than the experimental ones.
Again, this points to the necessity of correlation corrections.\\
In order to check the influence of an explicit treatment of the $d$ 
shells for Zn and Cd, we determined
the cohesive energies and lattice constants of ZnS and CdS using the
large-core pseudopotentials and corresponding basis sets as described
above. For ZnS we reach 72\%, for CdS 73\% of the experimental cohesive
energy, while the lattice constants are overestimated 
by 1\% only. However, the better agreement with experiment is due to 
a spurious compensation of the (still missing) correlation effects;
with the metal $d$ shells incorporated into the pseudopotential, 
the closed-shell repulsion
of these shells on the valence electrons of the neighbouring
atoms becomes too weak, cf.\ Ref.\ \onlinecite{leining96} for a thorough 
discussion in a molecular context.
This does not mean, on the other hand, that large-core pseudopotentials should
not be applied at all for Zn and Cd. Since closed-shell repulsion is mainly a
Hartree-Fock effect, correlation contributions are expected to come out quite
similar as with the small-core pseudopotentials. This expectation is fully borne
out by test calculations to be discussed in the following section.

\section{Many-body Corrections}

\subsection{Treatment of correlation effects}

There are  two important contributions of electron correlation 
which we have to consider in our calculations. The largest piece of
the correlation energy is due to the correlated motion of
the valence electrons. We account for it by applying the method of 
local increments 
to be explained in the second part of this subsection.
Intershell outer-core--valence correlations have little influence
on the cohesive energy but are important for the calculation
of the lattice constants and the bulk moduli. They are efficiently
simulated via a 
core-polarization potential\cite{mueller84}.
 
\subsubsection{Core-polarization potential}

The implicit description of core electrons by means of pseudopotentials or the 
explicit freezing of closed core shells at the Hartree-Fock
level implies a neglect of static and dynamic core polarization.
The former effect being due to static electric fields
is zero, for symmetry reasons, in isotropic solids of the type
considered here. The latter part, however, is non-zero and is related 
to core-valence
correlations. Especially with closed outer-core $d$-shells
the influence of dynamic core polarization
on bond lengths is known to be significant \cite{mueller84}, from 
molecular calculations. We simulate this effect
by means of a core polarization potential (CPP),
which describes the charge-induced dipole interaction between
valence electrons and cores:
\begin{equation}
V_{\rm CPP}=-\sum_{\lambda}\frac{1}{2} \alpha_{\lambda}
\vec{f}_{\lambda}^2 ;
\end{equation}
Here $\alpha_{\lambda}$
is the dipole polarizability of core $\lambda$, and $\vec{f}_{\lambda}$
is the field at
site $\lambda$ generated by valence electrons and surrounding cores:
\begin{equation}
\vec{f}_{\lambda}=
\sum_i^{N_{\rm V}}
\frac{\vec{r}_{\lambda i}}{r_{\lambda i}^3}
\left( 1-e^{-\delta_{\lambda}r_{\lambda i}^2} \right) -
\sum_{\mu (\neq \lambda)}
\frac{\vec{r}_{\lambda \mu}}{r_{\lambda \mu}^3}
\left( 1-e^{-\delta_{\lambda}r_{\lambda \mu}^2} \right) ;
\end{equation}
$r_{\lambda i}$ is the distance between valence electron
$i$ and core $\lambda$, $N_{\rm V}$ is the number of valence
electrons, and $r_{\lambda \mu }$ the distance between two cores. 
The cut-off parameter $\delta_{\lambda}$ is necessary in order
to remove
the singularity of the dipole interaction at $r_{\lambda i}=0$.
We took the parameters $\alpha_{\lambda}$ and $\delta_{\lambda}$
from Ref.\ \onlinecite{igelmann87}, where the CPP was adjusted in atomic
calculations to the spectra of single-valence-electron ions.\\   
Adding $V_{\rm CPP}$ (eq. 1) to the valence Hamiltonian in 
large-core pseudopotential SCF 
calculations corresponds to correlated calculations with simultaneous 
(single) excitations from both core
and valence-shell orbitals.
Using
the CPP in calculations, where the valence electrons are correlated, 
the additional coupling between core-valence and valence correlations 
is accounted for.

\subsubsection{Method of increments}

We determine (valence) correlation effects in infinite systems using
an expansion in terms of local increments.
Here we only want to sketch the basic ideas and some important formulae.
A formal derivation and more details can be found in Ref.\
[\onlinecite{paulus95}].
The method relies
on localized orbital groups (labeled $I,J$) generated in a SCF 
reference calculation using a suitable localization criterion 
(such as that of Foster and 
Boys\cite{foster60}).
The orbital groups may comprise single localized bond orbitals, 
in the case of covalently bonded
solids, or all of the (modified) atomic orbitals which can be 
assigned to an ion 
in the case of ionic solids.
One-body correlation-energy increments $\epsilon_I$ are obtained by
correlating each group of the localized orbitals separately while keeping
the other ones inactive. In the present work we use the
coupled-cluster approach with single and double substitutions
(CCSD).  This yields a first approximation to the correlation energy
\begin{equation} 
E_{\mbox{corr}}^{(1)}=\sum_I \epsilon_I ,
\end{equation} 
which corresponds to the correlation energy of
independent ions or independent bonds.\\
In the next step we include correlations of
pairs of orbital groups. Only the non-additive part $\Delta\epsilon_{IJ}$ of
the two-body correlation energy $\epsilon_{IJ}$ is needed.
\begin{equation}
\Delta\epsilon_{IJ}=\epsilon_{IJ}-(\epsilon_I+\epsilon_J).
\end{equation} 
Higher order increments are defined analogously. For
the three-body increment, for example, one has 
\begin{equation} 
\Delta\epsilon_{IJK}= \epsilon_{IJK} -( \epsilon_I +  \epsilon_J +
\epsilon_K) -(\Delta \epsilon_{IJ}+\Delta \epsilon_{JK}+ \Delta
\epsilon_{IK}).  
\end{equation} 
The correlation energy of the solid
is finally obtained by adding up all the increments with appropriate
weight factors:  
\begin{equation} E_{\mbox{corr}}^{\mbox{solid}} =
\sum_I  \epsilon_I + \frac{1}{2!}\sum_{IJ \atop I \not= J} \Delta
\epsilon_{IJ}+ \frac{1}{3!}\sum_{IJK \atop I \not= J \not= K} 
\Delta\epsilon_{IJK} + ... .  
\end{equation} 
It is obvious that by
calculating higher and higher increments the exact correlation energy
within CCSD is determined. However, the procedure described above is only
useful if the incremental expansion is well convergent, i.e., if
increments up to, say, three-body increments are sufficient, and if increments
rapidly decrease with increasing distance between localized
orbital groups.
These conditions were shown to be well met in the case of
covalently bonded solids \cite{paulus95,paulus96}, with an expansion 
in terms of localized bond orbitals,
and for ionic compounds like alkali halides \cite{doll97b} or
various oxides \cite{doll95}, using orbital groups of ions as basic entities.
When applying the method of increments to II-VI semiconductors we have to check
which orbital partitioning is more suitable and efficient, from a physical
as well as a computational point of view. Since this question also touches 
that of cluster embedding,
to be dealt with in the next subsection, we shall postpone its
discussion to Sec.III C.

\subsection{Embedding procedure}

Due to the local character of dynamical correlations,
the increments defined in the previous section should be fairly local 
entities. We use this property to calculate these increments in finite
fragments of the solid. Thereby, an appropriate embedding 
procedure simulating the 
\begin{figure}
\psfig{figure=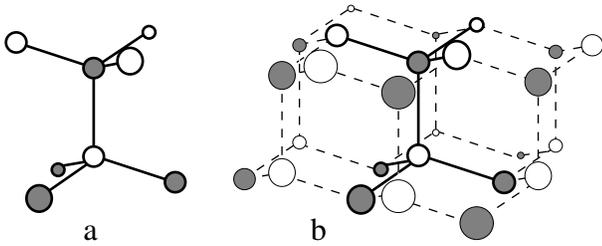,width=8cm,angle=-90}
\caption{\label{fig1}Figure a shows 
the $\rm{X}_4\rm{Y}_4$ cluster, where the white circles
indicate the group IIb elements, the shaded circles the
group VI elements; the hydrogen atoms are not drawn.
Figure b shows the $\rm{X}_{13}\rm{Y}_{13}$ cluster,
where the white/shaded circles connected by solid lines indicate
the group IIb/VI elements which are treated in the full basis set;
the hydrogen atoms are not drawn.}  
\end{figure}
influence of the infinite system surrounding the chosen cluster is of 
crucial importance.\\
For non-polar or only slightly polar semiconductors we use
a saturation of the dangling bonds with hydrogen atoms. As an example, 
a $\rm{X}_4\rm{Y}_4\rm{H}_{18}$ cluster is shown in Fig. \ref{fig1}a.
All the bond angles are chosen to be tetrahedral. The X---H and 
Y---H distances, respectively, are optimized
in CCSD calculations for a $\rm{XYH}_6$ cluster yielding
$d_{\rm ZnH}$=1.6643\AA, $d_{\rm CdH}$=1.7913\AA, $d_{\rm SH}$=1.3468\AA ,
$d_{\rm SeH}$=1.4658\AA\ and $d_{\rm TeH}$=1.6510\AA .
We performed test calculations for the II-VI semiconductors using
the $\rm{X}_4\rm{Y}_4\rm{H}_{18}$ cluster with the experimental XY 
distances of the solid. What
we find is a net charge
transfer of about 0.6 electrons (according to a Mulliken population analysis), 
from the hydrogen atoms bonded to group VI elements to
the hydrogen neighbours of the group IIb atoms.
This is mainly due to the electro-negativity of the group VI elements
being larger than that of hydrogen.
In order to avoid this charge transfer which generates an unphysical
charge distribution for the inner bonds, 
we tested a more sophisticated approach to hydrogen saturation.\\
The $\rm{X}_4\rm{Y}_4$ cluster is surrounded by an additional shell of X and Y
atoms at the sites of the bulk solid,
whose dangling bonds in turn are saturated with hydrogens 
(see Fig. \ref{fig1}b).
This yields a $\rm{X}_{13}\rm{Y}_{13}\rm{H}_{30}$ 
cluster which is still amenable to an {\em ab-initio} 
quantum-chemical treatment,
\begin{figure}
\psfig{figure=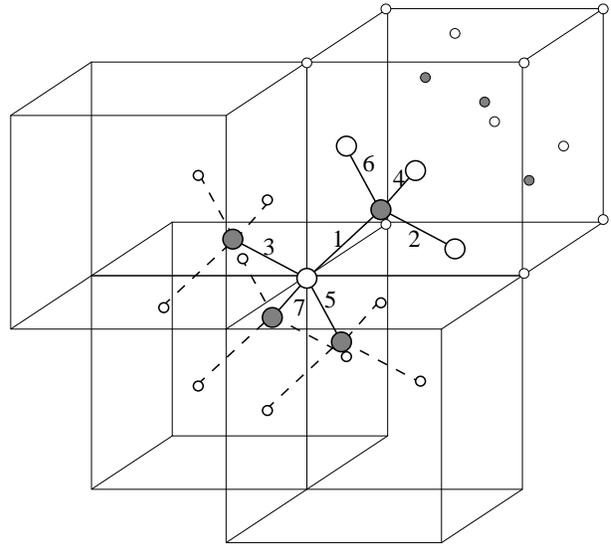,width=8cm}
\caption{\label{fig2}Figure 2 shows
the point charge embedding; in the center the $\rm{X}_4\rm{Y}_4$ cluster
treated with the full basis set (big circles connected by solid lines); the smaller circles
connected by dashed lines indicate the group IIb elements described with minimal
basis. In the upper right cube we show representatives for
the $4\times 4\times 4$ point-charge embedding 
(unconnected circles, white circle +2, shaded circle -2).}
\end{figure}
provided the additional X, Y atoms are treated with reduced accuracy 
(minimal $(1s1p)$ basis set).
Now, the charge transfer is much smaller than in the 
$\rm{X}_4\rm{Y}_4\rm{H}_{18}$ cluster, but a slight charge transfer of about 
0.2 electrons
still persists, from the upper plane where the group IIb elements 
are located, to 
the lower plane built up from the group VI elements. This could only be
avoided if the embedding in minimal-basis-set atoms were 
symmetrical with respect to
the inner $\rm{X}_4\rm{Y}_4$ fragment. This would result in a 
$\rm{X}_{22}\rm{Y}_{22}\rm{H}_{42}$ cluster, which is too large for our 
calculations.\\
Another possibility is the embedding with point charges. Here, one can
proceed to a very large surrounding, virtually without increasing the
computation time for the inner fragment. The $\rm{X}_4\rm{Y}_4$ cluster
is surrounded by point charges, with the group IIb and group VI atoms
simulated by point charges +2 and -2, respectively
(see Fig.\ \ref{fig2}). We extended the point-charge region to 
$4\times 4\times 4$ unit cells.
The point charges on the outer planes have the values $\pm$1, on the edges
$\pm\frac{1}{2}$ and on the corners $\pm\frac{1}{4}$.
(An analogous procedure works well for ionic compounds with 
NaCl-structure [\onlinecite{doll95,doll97b}].) \\
For the II-VI semiconductors there is a problem with the point-charge 
embedding: the electrons are too strongly localized at the outer 
group VI atoms.
In the infinite solid, there is
a charge transfer,
of about one
electron (according to Mulliken's population analysis), 
from each of the group VI ions 
to the group IIb neighbours.
In order to properly allow for this charge transfer in our cluster 
model, we replace
the point charges by 
large-core 
pseudopotentials, together  with  $(4s4p)/[1s1p]$ basis sets, for the 
group IIb neighbours
of the central X$_4$Y$_4$ unit.
(The $p$ function is necessary to generate the directional bonds.)
The modified point charge embedding for the $\rm{X}_4\rm{Y}_4$ cluster
is shown in Fig.\ref{fig2}.\\
We have tested all three embedding schemes just mentioned in calculations 
for Zn compounds. 
We calculated the one-bond increment for the
central bond of the X$_4$Y$_4$ unit, all two-bond increments up to 
second nearest neighbours,
and neighbouring three-bond increments;
the correlation energies obtained are
listed in Table \ref{embed}.
\begin{table}
\caption{\label{embed}Test calculations for the valence  correlation energy 
(in Hartree per unit cell) using different embedding schemes, cf.\ text.
The calculations are performed with basis A without the core-polarization
potential; third-nearest-neigbour bond contributions are neglected.}
\begin{tabular}{c||c|c|c|}
cluster model&ZnS&ZnSe&ZnTe\\
\hline\hline
X$_4$Y$_4$H$_{18}$&-0.1893&-0.1758&-0.1700\\
\hline
X$_{13}$Y$_{13}$H$_{30}$&-0.1847&-0.1650&-0.1489\\
\hline
X$_4$Y$_4$ + point charges&-0.1813&-0.1637&-0.1420\\
\end{tabular}
\end{table}
Although ZnS has the largest difference in electro-negativity,
the three embedding schemes yield very similar results.
For ZnTe, the simple saturation with hydrogen fails,
whereas the embedding in atoms with minimal basis and the embedding in
point charges provide similar results (error less than 5\%).
The correlation contributions obtained with the point-charge 
embedding are in all
cases smaller than those of the covalent cluster models, probably 
due to the more severe
restriction of the external space. Nevertheless, we favour
the point-charge embedding because of the higher computational efficiency. \\
We also tested 
the influence of the core-polarization potential on the lattice constant, 
for the three cluster models of ZnSe,
but found
no difference between the simple saturation with hydrogen atoms and the 
point-charge embedding. This shows that core-valence correlation is an
even more local entity than valence correlations.   
   
\subsection{Computational details}

\begin{figure}
\psfig{figure=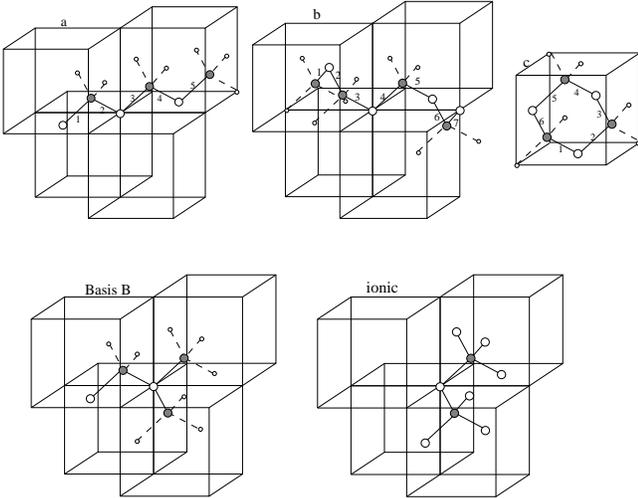,width=8.5cm}
\caption{\label{cluster}In the first line, we show
the clusters used for deriving third-nearest-neighbour increments; the second line
refers to calculations employing basis B and the $\rm{X}_7\rm{Y}_2$ cluster
used in the ionic approach, respectively. The notation and point-charge embedding is the same as in Figure 
\protect{\ref{fig2}}.}
\end{figure}

Calculating correlation-energy increments will be done in two steps.
First core-valence correlations are simulated in SCF calculations
with core-polarization potentials included for the 
two inner atoms of the $\rm{X}_4\rm{Y}_4$ cluster. Due to the local
character of core polarization (cf.\ Ref.\ [6b]), the incremental 
energy contributions from the
X and Y atoms are additive and directly transferable to the solid.
\\ The second step, i.e.\ the determination of
valence correlations and post-SCF core-polarization contributions
is more difficult. We first have to
decide which partitioning of the localized orbitals is appropriate
for the II-VI semiconductors. For the elementary and III-V semiconductors
it is natural to build up
the method of increments in terms of localized bonds, pairs and triples 
of such bonds; an accuracy of 1\% of
the correlation energy can be reached by carrying the correlation-energy 
expansion
up to third nearest 
neigbour two-bond increments and nearest-neighbour three-bond increments.
It seems reasonable to check whether such a scheme still works for the 
II-VI systems.
Bond increments up to second-nearest neighbours are accessible from
the $\rm{X}_4\rm{Y}_4$ cluster
shown in Fig. \ref{fig2}. For deriving increments involving third-nearest 
neighbours, we use the 
clusters shown in Fig. \ref{cluster}.
\begin{table} \caption{\label{inc}Correlation-energy increments for
ZnTe (in Hartree), determined at the CCSD level using basis set A. For
the numbering of the clusters and bonds involved, see Figs.
\protect{\ref{fig2}} and \protect{\ref{cluster}} . } 
\begin{tabular}{l|l|r|l} 
&cluster&Increment&Weight factor\\
\hline 
&\multicolumn{3}{c}{incremental expansion in terms of bonds}\\
\hline
$ \epsilon_I$&$\rm{X}_4\rm{Y}_4$/1&-0.0119037 &4\\
\hline 
$\Delta\epsilon_{IJ}$&$\rm{X}_4\rm{Y}_4$/1,2&-0.0082766&6\\
&$\rm{X}_4\rm{Y}_4$/1,3&-0.0017476&6\\
&$\rm{X}_4\rm{Y}_4$/2,3&-0.0003937&12\\ 
&$\rm{X}_4\rm{Y}_4$/2,5&-0.0003238&24\\
&Fig.\protect{\ref{cluster}}a/1,4&-0.0001562&6\\ 
&Fig.\protect{\ref{cluster}}a/2,5&-0.0000446&6\\ 
&Fig.\protect{\ref{cluster}}b/2,5&-0.0001166&24\\
&Fig.\protect{\ref{cluster}}b/3,6&-0.0000388&24\\
&Fig.\protect{\ref{cluster}}b/4,7&-0.0000984&12\\
&Fig.\protect{\ref{cluster}}b/1,4&-0.0000531&12\\
&Fig.\protect{\ref{cluster}}c/1,4&-0.0001490&12\\ 
\hline 
$\Delta \epsilon_{IJK}$&$\rm{X}_4\rm{Y}_4$/1,2,4&+0.007866&4\\ 
&$\rm{X}_4\rm{Y}_4$/1,3,5&+0.0001350&4\\ 
&$\rm{X}_4\rm{Y}_4$/1,2,3&-0.0000030&12\\ 
&$\rm{X}_4\rm{Y}_4$/1,2,5&+0.0000308&24\\
\hline
$E_{\rm corr}$&&-0.124408&\\
\hline
&\multicolumn{3}{c}{incremental expansion  in terms of ions}\\
\hline
$ \epsilon_I$&$\rm{X}_7\rm{Y}_2$&-0.097605&1\\
$\Delta\epsilon_{IJ}$&$\rm{X}_7\rm{Y}_2$&-0.004927&6\\ 
\hline
$E_{\rm corr}$&&-0.127170&\\
\end{tabular} 
\end{table}
As an example, we list the increments obtained for ZnTe, together  
with the appropriate
weight factors, in Table \ref{inc}.\\
The other possibility of orbital partitioning is the ionic approach
\cite{doll95}. Here, the localized orbitals are exclusively assigned to one of
the electro-negative group VI atoms.
The group defining a one-body increment comprises
the four $sp^3$-type bond orbitals connected to a given anion;
accordingly, the nearest-neighbour two-body increments are built up
from the orbital groups assigned to two neighbouring group VI atoms. 
The calculations are performed for
a  $\rm{X}_7\rm{Y}_2$ cluster (see Fig. \ref{cluster}). 
(A further extension is  computationally not feasible, at present:
inclusion of group VI atoms beyond nearest neighbours
would result in very large clusters, higher-order 
increments would involve correlation of  24 or more electrons. Thus, 
we cannot fully check 
the convergence of this approach.) The one-body and nearest-neighbour two-body
increment for ZnTe are listed in the second part of Table \ref{inc}.
Although the two approaches are not equivalent, neither with respect 
to the number of orbitals
simultaneously correlated nor with respect to the truncation 
criteria for the distance of correlated
orbitals,
the correlation
energies differ by 2\% only. This shows that we can bridge the gap between
incremental expansions appropriate for
covalent and ionic solids in an unequivocal way.
\\
In the calculations including valence correlations 
we used large-core pseudopotentials
for Zn and Cd combined with CPPs. In order to check the reliability of the 
large-core pseudopotentials for determining correlation contributions,
we performed test calculations for ZnS, ZnTe and CdS.
We calculated the one-bond increment in an
$\rm{XY}_4$ cluster using alternatively the 20-valence-electron and the 
2-valence-electron pseudopotentials for X = Zn, Cd. The absolute 
values of the increments obtained
differ by 5 mHartree, but the physically important differential 
correlation contributions
to the cohesive energy
differ by 1.5 mHartree at most, which is less than 1\% of
the cohesive energy. Therefore, for our purposes,
the large-core pseudopotentials can be considered as sufficiently accurate.
\\
Two different basis sets are used in our calculations.
Basis A is of valence-double-zeta quality. For the group VI elements we      
choose the $(4s5p)/[3s3p]$ valence basis set by Bergner et al. 
\cite{bergner93}, cf.\ also Sect.\ II. For Zn, we start from the 
$(4s2p)$ valence basis of Ref.\ \onlinecite{dolgunp}, add
the same inner $p$ functions as in the Hartree-Fock
calculation and end up (after contraction) with a $[3s3p]$ basis set
(cf.\ Table 2, numbers in parentheses).
For Cd, the atom-optimized primitive $(5s4p)$ basis set of Sect.\ 2
is contracted to
$[3s3p]$, too (cf.\ Table 2, numbers in parentheses). 
Furthermore, for each of the elements, we supply one $d$ polarization
function whose exponent is optimized in CCSD calculations for the free atom
(see Table \ref{polbas}, first column).\\
An enlarged basis B is generated by uncontracting the $sp$ valence basis sets
to $[4s4p]$ and replacing the single $d$ function by a $2d1f$ polarization set.
The exponents of the latter are energy-optimized in CCSD 
calculations for the free 
\begin{table} \caption{\label{polbas}Polarization functions used in
the CCSD calculations.} 
\begin{tabular}{c||c||ccc}
&basis A, 1d& \multicolumn{3}{c}{basis B, 2d1f}\\ 
\hline\hline
Zn&0.235&0.157& 0.330& 0.312\\ 
\hline 
Cd&0.203&0.163& 0.240& 0.268\\ 
\hline
S&0.479&0.269& 0.819 & 0.557\\ 
\hline 
Se&0.343&0.197& 0.462& 0.478\\ 
\hline
Te&0.221& 0.177 &0.281 & 0.360\\ 
\end{tabular} 
\end{table}
atoms (see Table \ref{polbas}). This basis is only applied to the
five largest increments which can be calculated in a ${\rm X}_{2}{\rm Y}_{3}$
cluster (see Fig.\ref{cluster}).\\ 
For hydrogen saturation, we use Dunning's
double-zeta basis \cite{dunning89} (without $p$ polarization function), or, in
the case of the ${\rm X}_{13}{\rm Y}_{13}$
clusters, its fully contracted (minimal basis set) version.

\section{Results and Discussion}

We applied the method of increments, as formulated in terms of 
localized bond orbitals, for 
calculating the core-valence (cv) and valence (vv) correlations  of
six II-VI compounds. As the first ground state property 
we calculated the cohesive 
energy at the experimental lattice constant (see Table \ref{coh}). At the 
Hartree-Fock level, where we use the small-core pseudopotentials
for Zn and Cd, we reach only between 60\% and 70\% of the experimental
value. Core-valence correlations have virtually no effect on the cohesive
energy; for the Zn-compounds they increase it by about 1\%, for
the Cd-compounds they yield a reduction of about the same percentage.
Valence correlations, on the other hand, have a significant influence 
on the cohesive energy.
With basis set A, the cohesive energy is increased
by about 20\%, with basis B even by about 30\%. This corresponds to 
about 60\% and 85\%, respectively, of the correlation contribution to the
cohesive energy. On the average, we recover 97\% of the experimental 
cohesive energy
\begin{table*} \caption{\label{coh}Cohesive energies per unit cell
(in Hartree), at different theoretical levels (cf.\ text); the 
Hartree-Fock values are obtained with small-core pseudopotentials 
for Zn and Cd, whereas the correlation calculations are performed 
with large-core pseudopotentials. Deviations
from experimental values\protect{\cite{harrison89}} 
(in percent) are given in parentheses.}
\begin{tabular}{c||cc|cc|cc|cc|c}
&\multicolumn{2}{c|}{HF}&\multicolumn{2}{c|}{HF+cv}&
\multicolumn{2}{c|}{HF+cv+vv}&\multicolumn{2}{c|}{HF+cv+vv}&exp\\
&&&&&\multicolumn{2}{c|}{basis A}&\multicolumn{2}{c|}{basis
B}&\\ 
\hline 
ZnS&0.159& (68\%)&0.160&(68\%)&0.204&(87\%)&0.224&(95\%)&0236\\
ZnSe&0.135&(69\%)&0.136&(70\%)&0.183&(94\%)&0.196&(100\%)&0.196\\
ZnTe&0.115&(64\%)&0.117&(65\%)&0.155&(86\%)&0.174&(97\%)&0.180\\
CdS&0.129& (62\%)&0.127&(61\%)&0.168&(80\%)&0.187&(89\%)&0.211\\
CdSe&0.110&&0.109&&0.151&&0.166&&\\
CdTe&0.099&(60\%)&0.099&(60\%)&0.132&(80\%)&0.153&(93\%)&0.164
\end{tabular} 
\end{table*}
for the Zn-compounds. For the 
Cd-compounds, the agreement is less perfect (about 91\%); it 
may be possible that an extended
basis for Cd,
including $g$ functions, yields further improvement. Errors due to 
the truncation of the 
incremental expansion can be estimated to
about $\pm$3\%. According to Harrison
\cite {harrison89}, the experimental error of the cohesive energy,
due to measuring the heat of
formation and the heat of atomization at different temperatures, is $\sim$1\%.
A comparison of our results with density-functional ones from 
literature is only possible for ZnSe, to the
best of our knowledge; the LDA value reported in Ref.\ [23] is
0.212 Hartree which corresponds to an overestimation of 8\%.
The percentage of the cohesive energy reached in our present calculations
is comparable to that for the
elementary and III-V semiconductors\cite{paulus95,paulus96} as well as 
to that for the ionic
systems\cite{doll95,doll97b}, showing that our approach
works well also for intermediate systems.\\
As in the case of cohesive energies, Hartree-Fock lattice 
constants are far from the experimental values, with deviations of up to 4.5\%.
In this case, however, the core-valence effects are decisive; they 
reduce the lattice constants
\begin{table*}
\caption{\label{latt}Lattice constants (in  \AA\ )
at different theoretical levels (cf.\ text); the
Hartree-Fock values are obtained with small-core pseudopotentials
for Zn and Cd, whereas the correlation calculations are performed
with large-core pseudopotentials. Deviations
from experimental values measured at room temperature  are given 
in parentheses.
\protect{\cite{landoltiii17b}}}
\begin{tabular}{c||cc|cc|cc|cc|c}
&\multicolumn{2}{c|}{HF}&\multicolumn{2}{c|}{HF+cv}&
\multicolumn{2}{c|}{HF+cv+vv}&\multicolumn{2}{c|}{LDA}&exp\\
&&&&&\multicolumn{2}{c|}{basis A}&&&\\
\hline
ZnS&5.5908& (+3.3\%)&5.4512&(+0.8\%)&5.4355&(+0.5\%)&5.42$^{\rm b}$
&(+0.2\%)&5.4100\\
ZnSe&5.8849&(+3.8\%)&5.7386&(+1.3\%)&5.7290&(+1.1\%)&
5.7839$^{\rm a}$
&(+2.0\%)&5.6676\\
ZnTe&6.3752&(+4.4\%)&6.1982&(+1.5\%)&6.1827&(+1.3\%)&6.1279
$^{\rm a}$&(+0.4\%)&6.1037\\
CdS&6.0465&(+3.9\%)&5.8730&(+0.9\%)&5.8853&(+1.2\%)&5.85$^{\rm b}$&
(+0.6\%)&5.8180\\
CdSe&6.3254&(+4.5\%)&6.1315&(+1.3\%)&6.1583&(+1.7\%)&6.07
$^{\rm b}$&(+0.3\%)
&6.0520\\
CdTe&6.7746&(+4.4\%)&6.5434&(+0.9\%)&6.5640&(+1.2\%)&6.40
$^{\rm b}$&(-1.3\%)&6.4860$^1$
\end{tabular}
{\footnotesize $^1$ calculated from the wurtzite structure\\
$^{\rm a}$ Reference[\onlinecite{freytag94}]\\
$^{\rm b}$ Reference [\onlinecite{vogel96}]}
\end{table*}
by about 3\% (see Table \ref{latt}), whereas valence
correlations have only a small influence. The latter fact is related to 
the opposite trends
caused by inter- and intra-atomic correlations; as a result, 
valence correlations can lead to a net
increase {\em or} decrease of the lattice constant.
For the lighter compounds, basis A is sufficient to describe the shortening of 
the bonds caused by intra-atomic correlations, whereas for the heavier
compounds basis A increases the lattice constant and a decrease 
is observed only with the enlarged basis B
(for example in CdTe: 
$a_{\rm basis A}^{cv+vv}$=6.5640\AA\ $\ge a^{cv}$ 
and $a_{\rm basis B}^{cv+vv}$=5.5437\AA\ $\le a^{cv}$, whereas for ZnS 
$a_{\rm basis A}^{cv+vv}$=5.4360\AA\ $\le a^{cv}$ 
and $a_{\rm basis B}^{cv+vv}$=5.4275\AA\ $\le a^{cv}$). 
A comparison with LDA results shows that
we reach about the same level of accuracy as Vogel et al.\cite{vogel96}, 
who treated
correlation effects with a self-interaction-corrected (SIC) pseudopotential.\\ 
The bulk moduli (see Table \ref{bulk}) calculated from the
first and second derivative of the potential curve 
at the experimental lattice constant are too large at the Hartree-Fock level.
Due to the uncertainties of the Hartree-Fock ground-state
energy values (cf.\ Sect.\ II), it is unreasonable to add the correlation energies to the 
Hartree-Fock ones and perform a fitting of the full potential curve:
the influence of correlations on the bulk modulus would be partly suppressed
by the 'noise' of the Hartree-Fock values. Therefore we seperately fitted
the core-valence and the valence correlation-energy contributions
through a quadratic fit, calculated the first and second 
derivatives of these curves at the experimental lattice constant,
and added them to the corresponding Hartree-Fock derivatives.
For the core-valence correlations, we determined the energy contributions
in the whole Hartree-Fock region, with steps of 1\% in the lattice constant.
These differential energies have numerical errors
of less than $10^{-5}$ Hartree, which yields negligible errors for the
bulk moduli; the subsequent fitting introduces errors
of about $\pm$2\%.
The core-valence correlations reduce the bulk moduli significantly,
by about 20\% on the average,  mainly due to the 
instantaneous deformation of the cores in the field of the
valence electrons.
\begin{table*}
\caption{\label{bulk}Bulk moduli (in Mbar)
at different theoretical levels (cf.\ text); the
Hartree-Fock values are obtained with small-core pseudopotentials
for Zn and Cd, whereas the correlation calculations are performed
with large-core pseudopotentials. Deviations
from (averages of) experimental values \protect{\cite{landoltiii17b}}
are given in parentheses.}
\begin{tabular}{c||cc|cc|cc|cc|cc|c}
&\multicolumn{2}{c|}{HF, min}&\multicolumn{2}{c|}{HF, exp}&
\multicolumn{2}{c|}{HF+cv}&\multicolumn{2}{c|}{HF+cv+vv}&
\multicolumn{2}{c|}{LDA}&exp\\
\hline
ZnS&0.744& (-3\%)&0.924&(+21\%)&0.799&(+4\%)&0.770&(+1\%)&
0.81$^{\rm b}$&
(+5\%)&0.748...0.784\\
ZnSe&0.634&(0\%)&0.885&(+39\%)&0.766&(+20\%)&0.738&(+16\%)&0.605$^{\rm a}$
&(-5\%)
&0.624...0.647\\
ZnTe&0.434&(-22\%)&0.734&(+32\%)&0.629&(+13\%)&0.604&(+9\%)&0.501$^{\rm a}$
&(-3\%)&0.50...0.612\\
CdS&0.593&(-9\%)&0.911&(+40\%)&0.792&(+22\%)&0.795&(+22\%)&
0.70$^{\rm b}$&
(+14\%)&0.615...0.69\\
CdSe&0.406&(-29\%)&0.727&(+34\%)&0.615&(+7\%)&0.616&(+7\%)&
0.66$^{\rm b}$&
(+20\%)&0.55...0.60\\
CdTe&0.337&(-29\%)&0.604&(+27\%)&0.498&(+5\%)&0.498&(+5\%)&
0.52$^{\rm b}$&
(+15\%)&0.425...0.528
\end{tabular}
{\footnotesize $^{\rm a}$ Reference[\onlinecite{freytag94}]\\
$^{\rm b}$ Reference [\onlinecite{vogel96}]}
\end{table*}
The influence of the valence correlations, which are calculated
at five points around the experimental lattice constant, is much smaller.
With basis A, they reduce the bulk moduli of the 
Zn-compounds by about 4\%. For the Cd-compounds the reduction is below
1\% with basis A; for CdTe, where we also applied basis B, there
is an reduction of about 4\%. Overall, the calculated bulk moduli 
are about 10\% too large compared with mean experimental values.
The error bars in our calculations can be roughly estimated to $\pm$7\%, all in all;
if we define
experimental error bars through the spread of the various experimental values,
they are of the same order of magnitude as our theoretical ones.
LDA calculations 
reach the same level of accuracy.
The advantage of
our approach is that we can discuss the influence of 
various electron-correlation effects 
on the lattice constant and the bulk modulus at a microscopic level.\\
        
\section{Conclusion}

We have determined ground-state properties (cohesive energies, 
lattice constants,
bulk moduli) of six II-VI semiconductors, at various theoretical levels.
The Hartree-Fock results have been obtained using the solid-state 
program {\sc Crystal92}, explicitly treating the outer-core $d$ 
shell of Zn and Cd
with small-core pseudopotentials
and crystal-optimized basis sets. For the cohesive energies, we recover
between 60\% and 70\% of the experimental values. The lattice constants
are overestimated by up to 4.5\% and the bulk moduli, evaluated at the
experimental lattice constants, are too large.
Accounting for
core-valence correlations by means of core-polarization potentials 
has only a slight
influence on cohesive energies but yields a large reduction of 
the lattice constants and the bulk moduli.
Correlations of the valence electrons are determined with the help 
of an incremental expansion
in terms of
localized bonds, and pairs and triples of such bonds; individual 
increments are evaluated at the coupled-cluster
(CCSD) level, for fragments
of the solid in a point-charge embedding.
Including valence correlations, we reach 85\% of the correlation
contributions to the cohesive energies, or $\sim$
95\% of the total experimental  values.
Valence correlations reduce the lattice constants only slightly,
leading to final values within 1\% of 
the experimental data; there is also moderate decrease of
the bulk moduli.\\
The present calculations show that a reliable microscopic description of
electron-correlation effects in II-VI semiconductors can be obtained in terms
of local excitations, and that a unified treatment 
(with uniform accuracy) is possible this way for
a variety of ionic as well as covalently bonded solids.

\end{document}